\documentclass[aps,prb,twocolumn,floatfix,superscriptaddress]{revtex4-1}
\usepackage[utf8]{inputenc}
\usepackage{graphicx}
\usepackage{epstopdf}
\usepackage{color}
\usepackage{xcolor,colortbl}
\usepackage{amsmath}
\usepackage{comment}
\usepackage{xr}

\newcommand{\bea}{\begin{eqnarray*}}
\newcommand{\eea}{\end{eqnarray*}}
\newcommand{\bne}{\begin{equation*}}
\newcommand{\ede}{\end{equation*}}

\newcommand{\bnen}{\begin{equation}}
\newcommand{\eden}{\end{equation}}
\newcommand{\bean}{\begin{eqnarray}}
\newcommand{\eean}{\end{eqnarray}}
\newcommand{\bnsn}{\begin{subequations}}
\newcommand{\edsn}{\end{subequations}}

\begin{document}
\title{Spectrally stable defect qubits with no inversion symmetry for robust spin-to-photon interface}
\author{P\'eter Udvarhelyi}
\affiliation{Department of Biological Physics, Lor\'and E\"otv\"os University, 
P\'azm\'any P\'eter s\'et\'any 1/A, H-1117 Budapest, Hungary}
\affiliation{Wigner Research Centre for Physics, Hungarian Academy of Sciences, P.O. Box 49, H-1525 Budapest, Hungary}
\author{Roland Nagy}
\affiliation{Institute of Physics, University of Stuttgart and Institute for Quantum Science and Technology IQST, Germany}
\author{Florian Kaiser}
\affiliation{Institute of Physics, University of Stuttgart and Institute for Quantum Science and Technology IQST, Germany}
\author{Sang-Yun Lee}
\affiliation{Center for Quantum Information, Korea Institute of Science and Technology, Seoul, 02792, Republic of Korea}
\author{J\"org Wrachtrup}
\affiliation{Institute of Physics, University of Stuttgart and Institute for Quantum Science and Technology IQST, Germany}
\author{Adam Gali}
\affiliation{Wigner Research Centre for Physics, Hungarian Academy of Sciences, P.O. Box 49, H-1525 Budapest, Hungary}
\affiliation{Department of Atomic Physics, Budapest University of Technology and Economics, Budafoki \'ut 8., H-1111 Budapest, Hungary}
\date{\today}

\begin{abstract} 
Scalable spin-to-photon interfaces require quantum emitters with strong optical transition dipole moment and low coupling to phonons and stray electric fields. It is known that particularly for coupling to stray electric fields, these conditions can be simultaneously satisfied for emitters that show inversion symmetry.
Here, we show that inversion symmetry is not a prerequisite criterion for a spectrally stable quantum emitter. We find that identical electron density in ground and excited states can eliminate the coupling to the stray electric fields.
Further, a strong optical transition dipole moment is achieved in systems with altering sign of the ground and excited wavefunctions.
We use density functional perturbation theory to investigate an optical center that lacks of inversion symmetry. Our results show that this system close to ideally satisfies the criteria for an ideal quantum emitter.
Our study opens a novel rationale in seeking promising materials and point defects towards the realization of robust spin-to-photon interfaces.
\end{abstract}

\maketitle

\section{Introduction}

Solid state defect quantum emitters are at the heart of quantum technologies. Quantum information technologies~\cite{Kane1998,Togan2010, Atature2018} and nanoscale sensor applications~\cite{RevModPhys.89.035002, Casola2018, PhysRevApplied.6.034001} require defects with long coherence times. Further, a quantum emitter with good optical properties allows for convenient optically-assisted spin state initialization and readout~\cite{Robledo2011,PhysRevLett.113.263602}. However, inhomogeneities in the host crystal can lead to degradation of optical properties.
A significant contribution is usually spectral diffusion and inhomogeneous broadening caused by the Stark shift effect. This spectral diffusion refers to the broadening of the integrated photoluminenscence zero-phonon-line (ZPL) after repetitive measurements on a single center as stray electric fields contribute fluctuating Stark shift to the transition. The mechanism is that optical excitation induces charge fluctuations of parasitic defects, which influence the optical properties of the investigated system~\cite{PhysRevB.61.R5086, PhysRevLett.107.266403}. Quantum emitters for quantum communication technology should emit predominantly coherent photons at large rate. However, especially for solid state emitters, photons are emitted either between  purely electronic states (zero-phonon emission) or between states with phonons being excited. Photons of the latter emission are not coherent. Therefore, the ratio of the emitted photons with zero-phonon contribution to the total emission, i.e., the Debye-Waller factor of the quantum emitter also categorizes the quality of the quantum emitter. Deficiencies in the spin coherence time can be principally circumvented by dynamical decoupling schemes, which can extend the coherence times further at the expense of longer measurement times~\cite{Awschalom2018}, that is an inevitable technique in such host materials where no spin-free isotopes are available. The intensity of coherent emission of defects with low Debye-Waller factors can be significantly improved by designing and building optical cavities around the quantum emitter which is a technological challenge but principally doable. On the other hand, the minimal spectral diffusion criterion is, in particular, inherently bound to such defect qubit property that cannot be efficiently and systematically circumvented either by means of materials design or quantum optical control of the quantum bits.

As the elimination of the parasitic defects near the quantum bits is a too high challenge for materials scientists, a simple solution to the problem of spectral diffusion is utilizing quantum defects with inversion symmetry that \emph{ab ovo} do not couple to static electric stray fields~\cite{PhysRevLett.113.113602}. This condition is intimately connected to the crystal structure of the host material of the quantum emitter. This symmetry requirement excludes compound semiconductors or insulator crystals that \emph{per se} do not host defects with inversion symmetry, although advanced production and processing technologies exist for those platforms. However, a plethora of compound semiconductors or insulators are potential candidates for hosting quantum emitters, e.g., defects in silicon carbide (SiC) have favorable coherence times in naturally abundant hosts without any dynamical decoupling procedures such as neutral divacancies~\cite{PhysRevLett.96.055501, Koehl2011, Falk2013, PhysRevX.7.021046} and negatively charged silicon-vacancies~\cite{PhysRevB.66.235202, PhysRevLett.109.226402, Kraus2013, Kraus2014, PhysRevB.93.081207, Widmann2015}.

In this paper, we show that the inversion symmetry is not a prerequisite criterion for a spectrally stable defect quantum emitter. We demonstrate this principle by means of density functional perturbation theory calculations on the so-called V1 center, i.e., the negatively charged silicon-vacancy at the so-called hexagonal ($h$) site in the compound semiconductor 4H SiC.

\section{Computational methods}
We determined the coupling of optical excitation to the external electric fields by calculating the permanent dipole moments in the corresponding $^{4}A_{2}$ ground state and $^{4}A_{2}$ excited state of the negatively charged silicon vacancy (V1) h-site defect in 4H-SiC~\cite{Janzen2009, Ivady2017} and compared those to the corresponding $^{3}A$ ground state and $^{3}E$ excited state of the negatively charged nitrogen-vacancy center in diamond~\cite{Davies1976, Goss1996, Gali2008} using density functional theory (DFT). We also determined the radiative lifetime of V1 center in 4H SiC.

\subsection{Electronic structure calculation}
We applied DFT for electronic structure calculation and geometry relaxation, using the plane-wave Vienna Ab initio Simulation Package (\textsc{VASP})~\cite{VASP1,VASP2,VASP3,VASP4}. The core electrons were treated 
in the projector augmented-wave 
(PAW) formalism~\cite{paw}. The calculations were performed with $420~\text{eV}$ plane wave cutoff energy and with $\Gamma$ centered $2\times2\times2$ k-point mesh for the 4H SiC supercell, $420~\text{eV}$ plane wave cutoff energy and $\Gamma$-point for the diamond supercell, respectively. We applied spinpolarized PBE functional in these calculations~\cite{PBE}. 
The model of V1 center in bulk 4H SiC was constructed using a 432-atom hexagonal supercell whereas we used the 512-atom simple cubic supercell to model nitrogen-vacancy (NV) center in diamond. The excited state electronic structure and geometry were calculated by constraint occupation of states, or $\Delta$SCF method~\cite{Gali2009}.

\subsection{Permanent dipole moment calculation}
We calculated the permanent dipole moments in the ground and excited state. The difference in the dipole moments is associated with the coupling parameter of the electric fields and the optical transition. To calculate the permanent dipole moments of the corresponding states, we used the \textsc{VASP} implementation of both Born effective charge calculation using density functional perturbation theory~\cite{PhysRevB.73.045112} and the Berry phase theory of polarization~\cite{PhysRevB.47.1651, PhysRevB.48.4442, RevModPhys.66.899}. In a DFT calculation, one can define the change in macroscopic electronic polarization ($\mathbf{P}$) as an adiabatic change in the Kohn-Sham potential ($V_{\text{KS}}$)
\begin{align}\label{response}
\frac{\partial \mathbf{P}}{\partial \lambda}=&-\frac{ife\hbar}{\Omega m_{e}}\sum_{\mathbf{k}}\sum_{n=1}^{M}\sum_{m=M+1}^{\infty}\\\nonumber
&\frac{\left<\psi_{\mathbf{k}n}^{(\lambda)}\right|\hat{\mathbf{p}}\left|\psi_{\mathbf{k}m}^{(\lambda)}\right>\left<\psi_{\mathbf{k}m}^{(\lambda)}\right|\frac{\partial V_{\text{KS}}}{\partial\lambda}\left|\psi_{\mathbf{k}n}^{(\lambda)}\right>}
{\left(\epsilon_{\mathbf{k}n}^{(\lambda)}-\epsilon_{\mathbf{k}m}^{(\lambda)}\right)^{2}}+c.c.\text{,}
\end{align}
where $f$ is the occupation number, $e$ elemental charge, $m_{e}$ electron mass, $\Omega$ cell volume, $M$ number of occupied bands, $\vec{\hat{p}}$ momentum operator.
The total change of polarization can be calculated by integrating over the adiabatic parameter $\lambda$.
In a periodic gauge, where the wavefunctions ($u_{\mathbf{k}}$) are cell periodic and periodic in the reciprocal space, the expectation values in \eqref{response} can be expressed as
\begin{equation}\label{pelement}
\left<\psi_{\mathbf{k}n}^{(\lambda)}\right|\hat{\mathbf{p}}\left|\psi_{\mathbf{k}m}^{(\lambda)}\right>=
\frac{m_{e}}{\hbar}
\left<u_{\mathbf{k}n}^{(\lambda)}\right|
\left[\nabla_{\mathbf{k}},H_{\mathbf{k}}\right]
\left|u_{\mathbf{k}m}^{(\lambda)}\right>\text{,}
\end{equation}
\begin{equation}\label{Velement}
\left<\psi_{\mathbf{k}n}^{(\lambda)}\right|\frac{\partial V_{\text{KS}}}{\partial\lambda}\left|\psi_{\mathbf{k}m}^{(\lambda)}\right>=
\left<u_{\mathbf{k}n}^{(\lambda)}\right|
\left[\frac{\partial V_{\text{KS}}}{\partial\lambda},H_{\mathbf{k}}\right]
\left|u_{\mathbf{k}m}^{(\lambda)}\right>\text{,}
\end{equation}
where $H_{\mathbf{k}}=\frac{1}{2m_{e}}(-i\hbar\mathbf{\nabla}+\hbar \mathbf{k})^2+V_{\text{KS}}$ is the periodic Hamiltonian.
Substituting \eqref{pelement} and \eqref{Velement} into \eqref{response}, the only contribution is $\left<u_{\mathbf{k}n}^{(\lambda)}\right|\nabla_{\mathbf{k}}\left|u_{\mathbf{k}n}^{(\lambda)}\right>$, as $\left<u_{\mathbf{k}n}^{(\lambda)}\right|\frac{\partial}{\partial\lambda}\left|u_{\mathbf{k}n}^{(\lambda)}\right>$ is periodic in the reciprocal space. The permanent dipole moment takes a form similar to the Berry phase expression
\begin{equation}
\mathbf{P}=\frac{ife}{8\pi^{3}}\sum_{n=1}^{M}\int_{\text{BZ}}d\mathbf{k}\left<u_{\mathbf{k}n}\right|\nabla_{\mathbf{k}}\left|u_{\mathbf{k}n}\right>\text{.}
\end{equation}
Using density functional perturbation theory (DFPT), $\nabla_{\mathbf{k}}\left|u_{\mathbf{k}n}\right>$ can be calculated from the Sternheimer equations with similar self-consistent iterations as in the self-consistent field DFT
\begin{equation}
\left(H_{\mathbf{k}}-\epsilon_{\mathbf{k}n}\right)\nabla_{\mathbf{k}}\left|u_{\mathbf{k}n}\right>=-\frac{\partial \left(H_{\mathbf{k}}-\epsilon_{\mathbf{k}n}\right)}{\partial \mathbf{k}}\left|u_{\mathbf{k}n}\right>\text{.}
\end{equation}

\subsection{Radiative lifetime calculation}\label{lifetime}
We determined the radiative transition rate between the ground and excited $^{4}A_{2}$ states by calculating the energy dependent dielectric function $\epsilon_{r}(E)$. The spontaneous transition rate is given by the Einstein coefficient
\begin{equation}
A=\frac{n\omega^{3}\left|\mu\right|^2}{3\pi\epsilon_{0}\hbar c^{3}}\text{,}
\end{equation}
where $n$ is the refractive index, $\hbar\omega$ is the transition energy, $\mu$ is the optical transition dipole moment, $\epsilon_{0}$ is the vacuum permittivity, $c$ is the speed of light. $\mu$ is proportional to the integrated imaginary dielectric function ($I$) of the given transition
\begin{equation}
\left|\mu\right|^2=\frac{\epsilon_{0} V}{\pi}\int \operatorname{Im}\epsilon_{r}(E) dE=\frac{\epsilon_{0} VI}{\pi}\text{,}
\end{equation}
where $V$ is the volume of the supercell. Thus, the radiative lifetime can be given by
\begin{equation}
\label{eq:tau}
\tau_{\text{r}}=\frac{3\pi^{2}\hbar c^{3}}{n\omega^{3}VI}\text{.}
\end{equation}

In our particular implementation, we applied the following procedure and parameters. We fit a Lorentzian function to the first peak of $\operatorname{Im}\epsilon_{r}(E)$. The results are $I=0.389~\mathrm{eV}$ and $\hbar\omega=1.387~\mathrm{eV}$. Using the refractive index $n=2.6473$ of 4H SiC and the cell volume of $V=4.5346~\mathrm{nm}^{3}$, the radiative lifetime can be calculated using Eq.~\eqref{eq:tau}.

\section{Results}

\subsection{Microscopic model of an ideal defect quantum emitter}
An ideal quantum emitter should show no spectral diffusion. This property can be achieved if the electric dipole moment remains either be zero or unchanged during the optical excitation process between the ground ($\left|g\right>$) and excited ($\left|e\right>$) states
\begin{equation}\label{eq:electric}
\left|\left<e\right|\vec{r}\left|e\right>-\left<g\right|\vec{r}\left|g\right>\right|^2=0 \text{.}
\end{equation}
On the other hand, the optical transition rate should be large
\begin{equation}\label{eq:transition}
\left|\left<e\right|\vec{r}\left|g\right>\right|^2 > 10~\mathrm{Debye}^{2} \text{.}
\end{equation}
The expectation values in Eqs.~\eqref{eq:transition} are generally nonzero according to the selection rules of quantum mechanics. As vector operators have $P=-1$ parity, the wavefunctions in the integral must have different parity, in order to result in a nonzero scalar.

For color centers with inversion symmetry, Eq.~\eqref{eq:electric} is satisfied as the individual integrals are zero, where the wavefunctions have either gerade (even) or ungerade (odd) parity. The high optical transition rate can be achieved by large overlap between a gerade orbital and an ungerade orbital in these optical centers. 

The main point of the present paper is the following statement: the inversion symmetry is not an ultimate criterion in simultaneous fulfillment of these requirements as Eq.~\eqref{eq:electric} may be satisfied without the restriction that all the individual terms in Eq.~\eqref{eq:electric} should be set to zero. We show below that other types of optical centers may satisfy Eq.~\eqref{eq:electric}.

\begin{figure}
\includegraphics{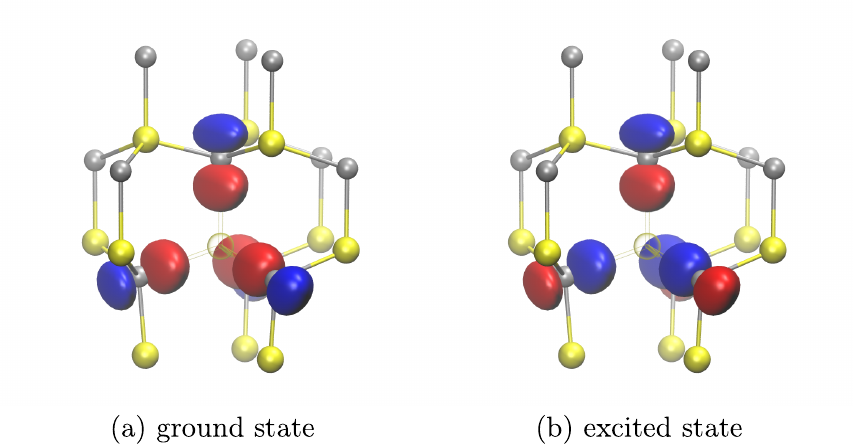}
\caption{Wavefunction of ideal defect quantum emitter with axial symmetry in a binary semiconductor. The different types of atoms are colored by gray and yellow balls, whereas the positive (negative) isovalues of the corresponding wavefunctions participating in the optical transition are depicted as red (blue) lobes.}
\label{fig:ideal}
\end{figure}

Defects usually have axial symmetry in compound semiconductors. In this case, Eq.~\eqref{eq:electric} can be satisfied for identical charge densities of the ground and excited state. In systems with axial symmetry, Eq.~\eqref{eq:transition} can be separated into two parts,
\begin{equation}
\left|\left<e\right|\vec{r}_\perp\left|g\right>\right|^2+\left|\left<e\right|\vec{r}_\parallel\left|g\right>\right|^2\text{,}
\end{equation}
where the first contribution is typically zero for orbitally non-degenerate ground state. The second contribution can be maximized by a large overlap of the wavefunctions (already satisfied by the same density requirement), if they are well separated in their signs along the symmetry axis, i.e., alternating phase of wavefunctions. This condition restricts the optical polarization to be parallel to the symmetry axis. A possible realization of the wavefunctions that fulfills the requirements detailed above is depicted in Fig.~\ref{fig:ideal} for a defect with axial ($C_{3v}$) symmetry.

The other requirement for an ideal quantum emitter is a high Debye-Waller factor, in order to achieve emission of coherent photons at high rate. This can be fulfilled for those defects that have negligible geometry relaxation upon optical excitation. 

\subsection{V1 center as an nearly ideal defect quantum emitter}
In this paper, we identify the $h$-site silicon-vacancy defect (V1 center) in 4H SiC as a nearly ideal quantum emitter with no inversion symmetry. Long coherence time was already reported for this defect making it a promising candidate for spin-based quantum applications~\cite{Simin2017, PRA2018}. In order to demonstrate that this defect also possesses all the optical requirements for quantum communication, we performed DFT calculations and compared this defect to the negatively charged nitrogen-vacancy (NV) center in diamond, which is known to exhibit a few tenths of GHz spectral diffusion even in high purity diamond samples~\cite{Siyushev:PRL2013}.

The results of Berry phase evaluation for macroscopic dipole moment calculation for V1 center in 4H SiC and NV center in diamond are shown in Table~\ref{tab:Berry_h_PBE}. We find that the change in the permanent dipole moments upon optical excitation for V1 center in 4H SiC is nearly 20 times smaller than that for NV center in diamond. This translates to weak coupling of optical excitation to stray electric fields for isolated Si-vacancies in 4H SiC, in good agreement with very recent experimental data~\cite{arxiv:1810.10296, arxiv:1811.01293}. 

\begin{table}
\caption{Macroscopic electric dipole moment of the negatively charged $h$-site Si-vacancy defect (V1) in 4H SiC and nitrogen-vacancy (NV) center in diamond as calculated within Berry phase approximation. ex(gr) notes the excited state electron configuration calculated with fixed ground state geometry.}
\begin{ruledtabular}
\begin{tabular}{lllll}
centre & transition & $\Delta p_{\text{ion}}~(e\AA)$ & $\Delta p_{\text{el}}~(e\AA)$ & $\Delta p_{\text{tot}}~(e\AA)$\\\hline
V1 & gr $\rightarrow$ ex & $0$ & $0.044$ & $0.044$ \\
V1 & gr $\rightarrow$ ex(gr) & $0$ & $0.039$ & $0.039$ \\
NV & gr $\rightarrow$ ex & $0.061$ & $0.842$ & $0.903$\\
\end{tabular}
\end{ruledtabular}
\label{tab:Berry_h_PBE}
\end{table}

\begin{figure}
\centering
\includegraphics{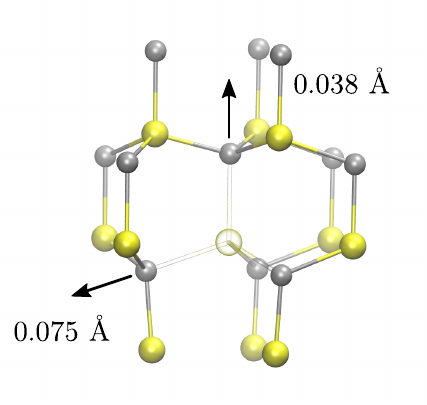}
\caption{Geometry relaxation of V1 center upon optical excitation. Small gray, large yellow and glass balls represent carbon, silicon atoms, and the vacant site, respectively. The movement of the first neighbor C-atoms are shown by arrows with the corresponding distances.}
\label{fig:relax}
\end{figure}

The geometry relaxation in the excitation transition of the V1 center is depicted in Fig.~\ref{fig:relax} which shows that the ions move outward going from the electronic ground state to the excited state. This leads to smaller than unity Debye-Waller factor in the luminescence spectrum. It is experimentally verified~\cite{PRA2018} to remain $\sim$0.5, which is about an order of magnitude larger than that of NV center in diamond~\cite{Alkauskas2014, Thiering2017, Atature2018}. The second row in Table~\ref{tab:Berry_h_PBE} shows the change in the dipole moments without relaxation effect at fixed ground state geometry. We conclude that the outward relaxation of ions upon optical excitation has little effect on the final difference in the dipole moments, and the change is associated with the nature of the ground state and excited state wavefunctions.


\subsection{Origin of permanent dipole moments and strong emission from V1 center in 4H SiC}
\begin{figure}
\includegraphics{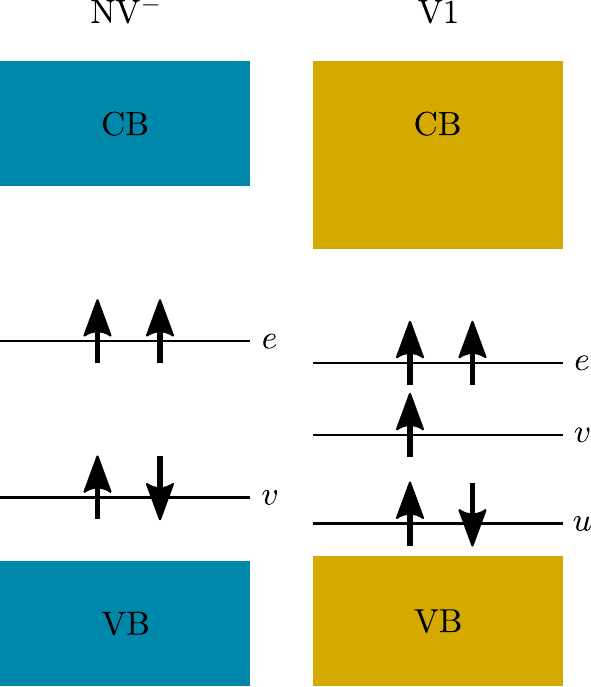}
\caption{Schematic visualization of ground state Kohn-Sham levels of NV and V1 center in the band gap, respectively. $v$ and $u$ states have non-degenerate $a_1$ symmetry whereas $e$ state is double degenerate. The occupation of these levels shows the ground state electronic configuration. In the spin minority channel (spin down), the electron is promoted to the $e$ level in NV center and $v$ level in V1 center, respectively, in the excited state electronic configuration.}
\label{fig:levels}
\end{figure}

The results on the microscopic level can be interpreted by considering the electron density of defect states in the ground state. The corresponding in-gap defect levels and labels are shown in Fig.~\ref{fig:levels}. This approximation neglects relaxation effects of ions and the effect of the delocalized electron bath. We approximate the origin of total change of electric dipole moment by studying the difference of electron density of these in-gap Kohn-Sham states, which change occupation during optical transition (V1 $u$ and $v$ levels and NV $v$ and $e$ levels in Fig.~\ref{fig:levels}).

\begin{figure*}
\includegraphics{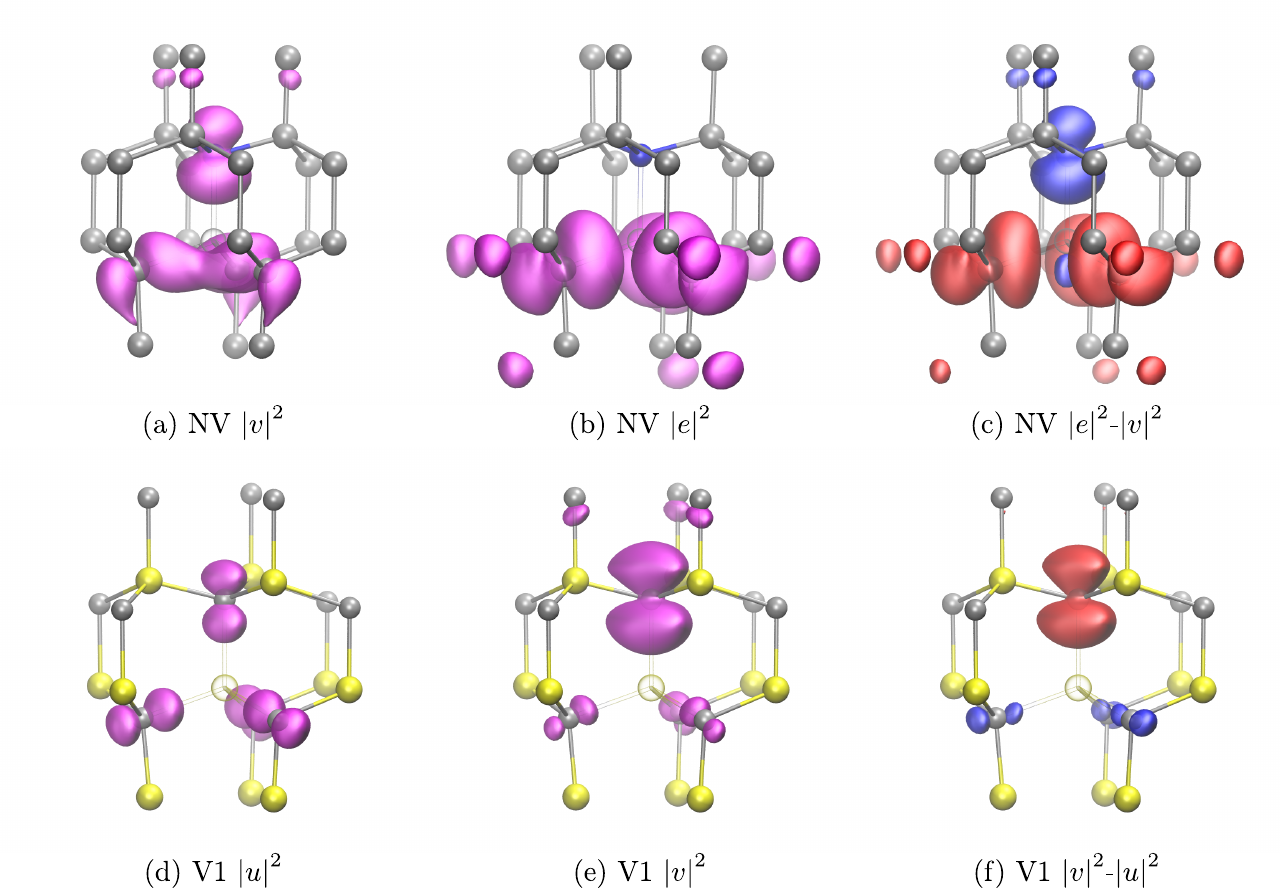}
\caption{Electron density isosurfaces (0.025\AA $^{-3}$ in purple color) of the ground state in the minority spin channel for the localized Kohn-Sham states of NV center in diamond (a, b) and V1 center in 4H SiC (d, e). Their positive (negative)  difference is shown in red (blue) colors in (c, f). Gray, yellow and blue balls represent carbon, silicon and nitrogen atoms, respectively. Vacancy is represented by a glass ball.}
\label{fig:densities}
\end{figure*}

To visualize this scenario, we plot the electron density in the minority spin channel of the localized Kohn-Sham orbitals in the ground state (Fig.~\ref{fig:densities}) which represents the ground state and excited state in the photoexcitation process. NV in diamond shows a rather large change in localization and magnitude of the electron densities going from the excited state to the ground state which leads to a considerable change of electric dipole moment. However, V1 center manifests small change in localization and magnitude of the electron densities that suggests small change of electric dipole moment. 

\begin{figure}
\centering
\includegraphics{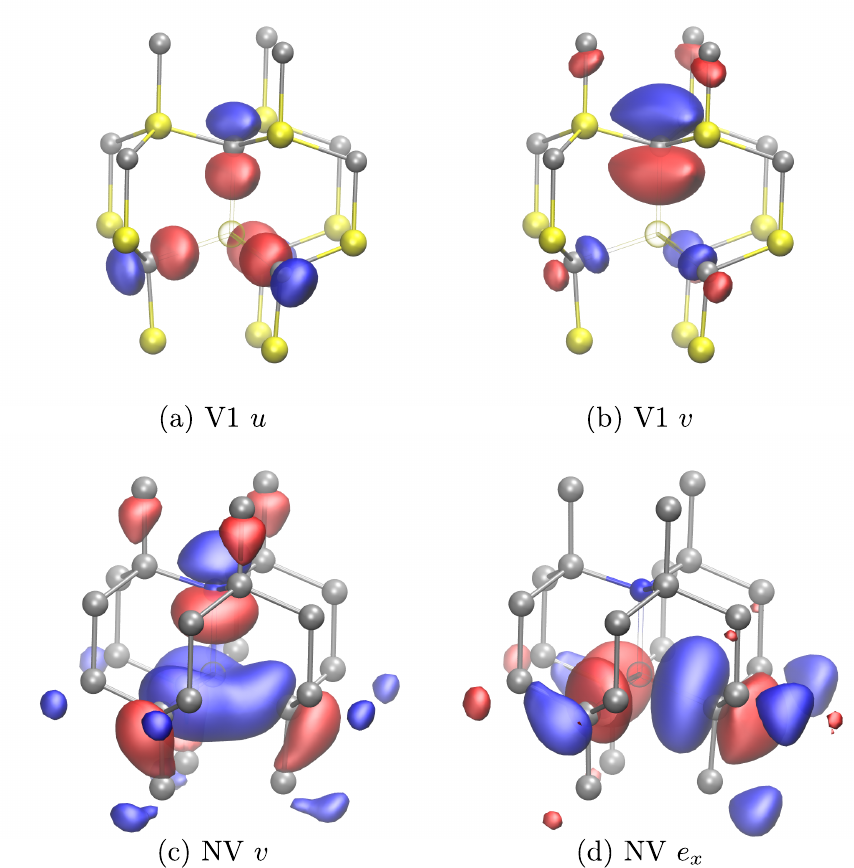}
\caption{Square moduli of the corresponding wavefunctions of the ground state in the minority spin channel in V1 (a, b) and NV (c, d) center. The same color code is used as in Fig.~\ref{fig:densities}(f).}
\label{fig:wavefunction}
\end{figure}

Figure \ref{fig:wavefunction} (a, b) shows overlap of same phase for the lobe in axial direction resulting in a positive contribution. The basal lobes overlap with different phase resulting in negative contribution. These spatially well-separated contributions with opposite sign integrate to a rather large transition dipole moment. The same argument can be made for NV center in the perpendicular direction to the defect axis. The calculated radiative lifetime of V1 center is 12~ns which is fairly comparable to the calculated 11~ns~\cite{Siyushev2013} and measured $\sim$12~ns~\cite{Batalov2008} lifetime of NV center in diamond.

These lines of the discussion are inspired by Eqs.~\eqref{eq:electric} and \eqref{eq:transition} that indeed explain well our \emph{ab initio} results on the change of the permanent dipole moments versus the radiative lifetime.


\section{Discussion}
Our findings have important implications in seeking materials hosting defect quantum emitters for realizing robust spin-to-photon interfaces. It is a common wisdom in the present literature that only defects with inversion symmetry are decoupled from stray electric fields, which significantly constrains the type of host materials. For instance, only 84 centrosymmetrical three-dimensional (3D) crystal structures exist among the total 230 feasible 3D crystal structures. In particular, commercially available wide band gap semiconductors, such as SiC, ZnO or GaN are all not centrosymmetrical. In addition, two-dimensional (2D) transition-metal dichalgonides (TMD) are also \emph{per se} not centrosymmetrical, thus they cannot host single photon defect emitters with inversion symmetry. We show theoretical arguments that the quest of inversion can be relaxed and other types of defect quantum emitters are in contention, such as V1 center in 4H SiC. In particular, 4H SiC hosts spin-active emitters, e.g., a recent study on molybdenum~\cite{Csore2016, Bosma2018}, that might have also favorable optical stability against stray electric fields. Quantum emitters in other semiconductors such as ZnO~\cite{Morfa2012, Neitzke2015} and GaN~\cite{Berhane2017, Zhou2018} should be also revisited in this regard. Our study also provides a guidance to design novel defect quantum emitters in 2D compound materials~\cite{Tran2016, Tongay2013, Saigal2016, He2015, Srivastava2015} for stable spin-to-photon interface at a single photon level.  

\section*{Acknowledgment} The support from National Research, Development and Innovation Office in Hungary (NKFIH) Grant Nos.\ 2017-1.2.1-NKP-2017-00001 (National Quantum Technology Program) and NVKP\_16-1-2016-0043 (NVKP Program) as well as Grant No.\ NN127902 (EU QuantERA Nanospin consortial project), and from the EU Commission (ASTERIQS project with Grant No.~820394) as well as BMBF project Q.Link.X is acknowledged. We acknowledge the discussion with Alp Sipahigil.




%

\end{document}